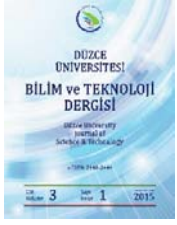

# Düzce Üniversitesi
# Bilim ve Teknoloji Dergisi

*Araştırma Makalesi*

# Teknokent'lerde Geliştirilen Yazılım Projelerinin Risk Analizi ve Başarı Düzeyleri

M. Hanefi CALP [a,*], M. Ali AKCAYOL [b]

[a]*Bilişim Enstitüsü, Gazi Üniversitesi, Ankara, TÜRKİYE*
[b]*Bilgisayar Mühendisliği, Mühendislik Fakültesi, Gazi Üniversitesi, Ankara, TÜRKİYE*
* *Sorumlu yazarın e-posta adresi: mhcalp@gazi.edu.tr*

### ÖZET

Günlük hayatın hemen her alanında ihtiyaç duyulan yazılım projeleri her geçen gün daha fazla büyümekte ve daha çok önem arz etmektedir. Bununla birlikte, büyüyen bu yazılımlar beraberinde daha karmaşık bir yapıyı doğurmaktadır. Bu bağlamda, yazılım projelerinin analiz ve kontrol süreci, ayrıca başarı düzeyleri merak edilmektedir. Dolayısıyla bu çalışmada, geliştirilen anket formlarında bulunan kontrol listeleri aracılığıyla, Teknokent'lerde geliştirilen yazılım projelerinde karşılaşılan risk faktörlerini belirlemek, analiz etmek ve bu projelerin başarı düzeylerini ortaya koymak amaçlanmıştır. Bu amaçla, gerekli olan veriler toplanmış, yazılım risk faktörleri belirlenmiş ve bu faktörler analiz edilmiştir. Araştırma kapsamında elde edilen verilerin analizi için betimsel analiz yönteminden yararlanılmış olup, Microsoft Excel 2010 ve IBM SPSS Statistics 21.0 programları kullanılmıştır. Ayrıca, çalışmada analiz sonuçlarına ve geliştirilen yazılımların başarı düzeylerine ayrıntılı bir şekilde yer verilmiştir. Araştırma sonuçlarına göre özellikle; proje süresi, bütçesi, personel sayısı ve hedeflerdeki sapmalara rağmen Teknokent'lerde geliştirilen yazılım projelerinin yaklaşık %95'inin başarıyla sonuçlandırılmış ve müşteriye teslim edilmiş olduğu gözlenmiştir.

**Anahtar Kelimeler:** *Yazılım, Risk faktörleri, Risk analizi, Proje başarısı*

# Risk Analysis and Success Levels of the Software Project Developed in Technocity

### ABSTRACT

Software projects needed on almost every aspect of daily life, are growing and gaining more importance more and more every day. However, this software that growing raises a more complex structure. However, this software that growing is a more complex structure. In this context, analysis and control process and success levels of the software projects are wondered. Therefore, in this study, through the checklist in questionnaire developed, found in, is aimed to identify and analyze risk factors which encountered in the software project that performed in Technocity and to determine the levels the success of this project. For this purpose, the necessary data were collected, were identified software risk factors and this factors were analyzed. The descriptive analysis method was used for the analysis of the data obtained within research and Microsoft Excel 2010 and IBM SPSS Statistics 21.0 software. In addition, are included to analysis results and the level of success of the developed software in the study. According to results of research, particularly; was observed to be completed successfully the about 95% of software projects developed in Technopolis and delivered to the customer in spite of divergence in project duration, budget, personnel and objectives.

# I. GİRİŞ

GÜNÜMÜZ endüstrileri arasında önemli bir yere sahip olan yazılım endüstrisi, geliştirilen ürünlerin büyüklüğü ve insan yaşamının her alanında yer alması ile bu önemini daha da arttırmaktadır. Yazılım endüstrisi tarafından üretilen yazılımlar; cep telefonlarından bilgisayarlara, vatandaşlık işlemlerinden sağlık sektörüne, askeriyeden enerjiye hem üretim hem de tüketim alanında kullanılmaktadır. Dolayısıyla, bu alanlardaki ihtiyaçların karşılanması için de projeler hazırlanmaktadır. Her geçen gün daha fazla yaygınlaşan ve vazgeçilmez hale gelen bu projelerin hatasız geliştirilmesi yazılım firmaları açısından önemlidir. Bunun için iyi kodlama yapmakla birlikte, sonucu önemli ölçüde etkileyecek risk analizi gibi süreçler başarıyla gerçekleştirilmelidir. Bunun için ise, bu konudaki standartların en iyi şekilde uygulanması ve daha önce yapılan hatalardan ders alınması gerekmektedir [1].

Yazılım projelerinin belirsiz olması, bu projelerin genellikle daha fazla risk içermesine ve daha fazla maliyete ihtiyaç duyulmasına sebep olmaktadır. Bu durum, yazılım projelerinde risk analizinin ne kadar önemli olduğunu; aynı zamanda, analiz sürecinin önceden eksiksiz bir şekilde uygulanmasının proje başarı düzeylerini ne kadar çok etkilediğini açıkça göstermektedir. IEEE'ye göre yazılım projelerinde projeye ait risklerin belirlenme oranı %90 olmasına rağmen, araştırmalarda risklerin belirlenme oranının %50-70 arasında olduğu gözlemlenmiştir. Zaman ve harcanan iş gücü hesaplandığında risklerin ve sonuçlarının önceden belirlenip, risklerin önem sırası ve olasılıklarına göre önlemler alınması şarttır [2,3]. Charatte (1996) ise, büyük ölçekli yazılım projelerinde risk analiz süreçlerinin daha zor geçtiğini belirtmiş ve bu işlemin proje aşamasındaki önemini vurgulamıştır [4].

2004 yılında, Standish Group International'ın üzerinde çalıştığı projelerin %53'ünün gecikmiş ya da bütçeyi aşmış, %18'inin terk edilmiş, geri ölçekli ya da değiştirilmiş olduğu ortaya konmuş olup, bu projelerden yalnızca %29'unun zamanında ve bütçeye uygun olarak tamamlandığı belirtilmiştir. Yazılımların, bütçeyi aşması ve teslim süresinde gecikme olması; büyük oranda yönetim ve analizlerle ilgilidir. Dolayısıyla, yazılımlarda oluşabilecek risklerin önceden belirlenmesi, analizlerin eksiksiz bir şekilde uygulanması ve yönetilmesi gerekmektedir [5].

Yazılım projelerinin başarı düzeyleri açısından risk analizi ve yönetiminin önemli olduğu konusunda bilim adamları hemfikir olup, özellikle şu kuralların önemine dikkat çekmektedirler.

- Risk, ne kadar erken bulunursa, o kadar erken azaltılır. Böylece, daha düşük geliştirme maliyeti ile yönetilir.

- Potansiyel tehlike kontrol edilerek ve azaltılarak, çeşitli risklerin reaktif olması önlenmiş olunur. Bu yolla, gelecekte oluşabilecek birçok ciddi problemden sakınılabilir.

- Uygun bir şekilde tanımlanmış risk yönetim süreç modeli için, çeşitli risk türlerinin öncelik sırasına koyulması gerekmektedir.

Bu çalışmada, geliştirilen anket formlarında bulunan kontrol listeleri aracılığıyla, Teknokent'lerde gerçekleştirilen yazılım projelerinde karşılaşılan risk faktörlerini belirlemek, analiz etmek ve bu projelerin başarı düzeylerini ortaya koymak amaçlanmıştır. Çalışmanın ikinci bölümünde, yazılım projelerinde risk analizi konusuna; üçüncü bölümünde, yapılan çalışmanın ayrıntılarına (yöntem, araştırmanın önemi ve amacı, evren ve örneklem, verilerin toplanması ve güvenirlik analizi); dördüncü bölümde, bulgular ve yorumlara ve son olarak beşinci bölümde ise, çalışmadan çıkarılan sonuç ve önerilere yer verilmiştir.

## II. YAZILIM PROJELERİNDE RİSK ANALİZİ

Araştırmalara göre, bilinen birçok risk analizi yöntemi vardır. Yazılım mühendisliğinde risk analizi, bir projedeki yüksek riskli elemanları tanımlamak için kullanılır. Bu faaliyet, risk azaltma stratejilerini dokümante ederek gerçekleşir. Risk analizinin, sistemdeki kritik durumları değerlendirmesi, yazılım tasarım evresinde önemli olduğunu göstermektedir [6,7]. Risk analizinin amacı, riskleri ve onları düzelttikten sonraki katkılarını daha iyi anlamaktır. Başarılı bir analiz; problem tanımlama, problem formülasyonu ve veri toplama gibi temel elemanları içermektedir [8]. Yazılım risk analizi, genellikle nesnel bir bakış açısından ziyade öznel bir bakış açısıyla gerçekleştirilmektedir. Yazılım Mühendisliği Enstitüsü ve diğerleri, riskleri temelde yüksek/orta/düşük olacak şekilde gruplayarak, bu risklerin üstesinden gelirler. Bu çalışmalar, modeller ve kurallar, risk altındaki ürünü daha iyi bir hale getirebilmek için ilk şarttır [9-13]. Risk analizi sürecinde, öncelikle herbir tanımlı risk ve etkileri, olasılıkları, şiddetleri bağımsız bir şekilde değerlendirilerek analiz edilir [14]. Analiz; *metrikler, karar ağaçları ve senaryo analizleri* gibi farklı teknikler kullanılarak yapılabilir [15]. Daha sonra riskler önceliklendirilir ve en öncelikli risk listesi oluşturulur [16]. Analiz sonuçlarına göre, her bir risk veya risk gruplarını yönetmek için bir plan önerilir. Son olarak, öncelikli risk listesi, onay için paydaşlarla görüşülür [17].

Sonuç olarak bu sürecin hedefi; risk senaryolarını hazırlamak ve risk kontrol eylemlerinin planlanabilmesi için proje risklerini detaylı olarak tanımlamaktır. Risk analizi süreci; risklerin olasılıklarını, etkilerini hesaplayabilmek ve önemli riskleri belirlemek için risk bileşenlerinin analiz edilmesi sürecidir. Süreç, ilk risk değerlendirmesinde ve her yeni tespit edilen risk için yeniden başlatılacaktır. Sürecin girdisi, risk listesidir. Çıktısı ise, önemli riskler konusunda sürece katılan personelin fikir birliği ile üretilen önceliklendirilmiş risk senaryolarıdır. Risk analizi sürecinde temel olarak iki alt süreç gerçekleştirilmektedir. Bunlardan birincisi, seçilen risklerin senaryolar halinde belgelendirildiği *"risk senaryosu geliştirme"*; ikincisi ise, risk senaryolarının sıralanarak *"önceliklendirilmesi"*dir.

### A. RİSK SENARYOSU GELİŞTİRME

Risk senaryosu geliştirme alt süreci, temel riskler için senaryolar geliştirmeyi amaçlayan ve bu maksatla analiz grafiği (Şekil 1) ve diğer çizim araçlarının kullanıldığı, süreç çıktısı olarak en muhtemel riskler için risk senaryolarının üretildiği ve proje yöneticisinin sorumlu olduğu bir süreçtir.

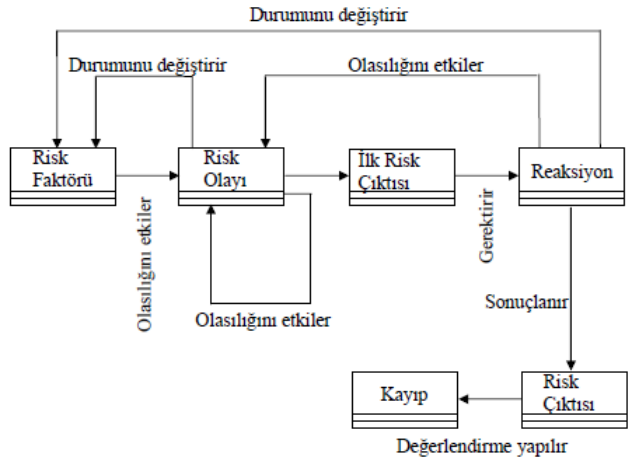

***Şekil 1.*** *Analiz grafiği [5]*

Risk maddelerinin seçilmesi, bir sonraki adım olan risk önceliklendirmeye de katkı sağlayacaktır. Tüm risk maddeleri gözden geçirilerek hangilerinin analiz sürecine dâhil edileceği açıkça belirtilmelidir. Bu konudaki kural, en önemli risklerin seçilerek senaryolaştırılması ve risk kontrol eylemleri gerçekleştirilinceye kadar risk grupları içinde kalan diğer risk maddeleri için senaryo geliştirmeye devam etmektir [5].

## *B. RİSK ÖNCELİKLENDİRME*

Risk yönetimi için ayrılan kaynakların sınırlı olması nedeniyle, tüm risk senaryolarının analiz edilmesi veya ortadan kaldırılması mümkün olamamaktadır. Bunun yerine önemli riskler üzerine odaklanmak, zaman ve kaynak ayırarak onları yönetmek daha uygundur. Bunun başarılması için risk senaryolarının önem derecelerine göre sıralanması gerekmektedir. Risk senaryolarının önceliklendirilmesi için, her bir riske göre fayda kaybının ve ihtimalinin hesaplanmasına ihtiyaç vardır. Bu iki tahmin problemi, çok çeşitli zorlukları da beraberinde getirmektedir. Prensipte geçmişe yönelik elde mevcut bilgilerin yetersiz olması ve çevrenin sürekli değişmesi hususlarına bağlı olarak, ihtimal hesabı yapmak çok zordur. Hatta bazen imkânsızdır. Fayda kaybının tahmin edilmesine ilişkin zorluk ise, dikkate alınması gereken faktörlerin çokluğundan ve her bir proje ilgilisi için faydanın hangi yapıda olduğunun tam olarak bilinmemesinden kaynaklanmaktadır. Önceliklendirme sürecinde, senaryoların olasılık ve fayda kaybı sıralamaları kullanılmakta ve senaryoların diğerlerine göre orantısal etkinlikleri araştırılmaktadır [5]. Risk analizinde; riskler belirlenir, risklerin etkileri ve oluşma olasılıkları tahmin edilir, riskler önemlerine göre sıralanır. Bu aşamadan sonra risklerin etki değerleri belirlenir. Bu etki değerleri ; örneğin, kritik, ciddi (program sonlanmayacağı fakat tarih ve bütçe bakımından artış ve ertelemelerin yaşanabileceği durumlar), orta seviye, en az (risk oluştuğu zaman, tarihte ve bütçede az bir artış oluyorsa) ve etkisiz (riskin hiçbir etki değeri bulunmuyorsa) olarak sınıflandırılır. Risklerin etki değer tanımları ve risk olasılık değerleri ile risk seviyesi hesabı yapılarak riskin tolere edilebilir olup olmadığına bakılır. Risk yönetimi aşamasında ise işlem sonlandırılabilir, tolere edilebilir veya transfer edilebilir [18].

## III. YÖNTEM

Bu bölümde; araştırmanın yöntemi, önemi ve amacı, verilerin toplanma aşamaları ve bu süreçte karşılaşılan sorunlar, evren ve örneklem, güvenilirlik ve frekans analizleri yer almaktadır. Bu bağlamda, çalışmada öncelikle uzmanlarla görüşülüp beyin fırtınası yapılarak "Takım Üyeleri (Geliştirici ve Test Elemanı)" ve "Yönetici (Analist, Uzman, Takım Lideri, Proje yöneticisi ve Genel Yönetici)" başlıkları altında iki adet anket formu hazırlanmıştır. Sözkonusu formlar, toplamda elli (50) adet soru olmak üzere üç bölümden oluşmaktadır. İlk bölümde, yazılım projelerinin; proje alanı, pozisyonu, proje süresi, proje bütçesi ve personel sayısı bilgileri; ikinci bölümde, proje süresinde, bütçesinde, personel sayısında ve hedeflerde sapmalar olup olmadığı, aynı zamanda projenin genel olarak amacının başarıyla gerçekleşip gerçekleşmediği; üçüncü bölümde ise, sekiz ana başlık altında toplanan ve 5'li likert yöntemiyle risk düzeyleri (çok düşük, düşük, orta, yüksek, çok yüksek) ve oluşma sıklıkları sorgulanan kırk (40) adet risk faktörü bulunmaktadır. Dolayısıyla bu çerçevede, geliştirilen anket formlarında bulunan kontrol listeleri aracılığıyla, gerekli olan veriler toplanmış, karşılaşılan risk faktörleri belirlenmiş ve sözkonusu faktörler analiz edilmiştir. Böylece, Teknokent'lerde geliştirilen yazılım projelerinde karşılaşılan riskler ve bu projelerin başarı düzeyleri ortaya konulmuştur. Çalışmada, Takım Üyesi ve Yönetici olarak hazırlanan iki adet formdan elde edilen veriler, betimsel analiz yöntemi ile analiz edilmiştir. Ayrıca, bu süreçte Microsoft Excel 2010 ve IBM SPSS Statistics 21.0 programlarından yararlanılmıştır.

*A. ARAŞTIRMANIN ÖNEMİ VE AMACI*

Geliştirilen yazılımların, hem üretim hem de tüketim alanı olmak üzere hemen hemen tüm alanlarda kullanılması, bu alanlardaki ihtiyaçların karşılanması için birçok farklı projelerin hazırlanmasına sebep olmaktadır. Bu projelerin hatasız geliştirilmesi için ise, yazılım risk analiz süreçlerinin en iyi şekilde uygulanması ve daha önceki projelerde yapılan hatalardan ders alınması gerekmektedir. Dolayısıyla bu noktada, yazılım projelerinde karşılaşılabilecek risk faktörlerinin doğru bir şekilde belirlenmesi, risk analizi ve yazılım projelerinin başarı düzeyleri çok önem kazanmaktadır.

Çalışmanın amacı ise, Teknokentlerde geliştirilen yazılım projelerinde karşılaşılan risk faktörlerini belirlemek, bu faktörleri analiz etmek ve sözkonusu projelerin başarı düzeylerini ortaya koymaktır. Böylece, proje yöneticileri açısından, yazılım risklerinin meydana gelmeden önlenmesi anlamına gelen proaktif (önleyici) risk stratejisi uygulamak daha kolay bir hal alacaktır.

*B. VERİ TOPLAMA SÜRECİNDE KARŞILAŞILAN SORUNLAR*

Araştırmanın veri toplama sürecinde karşılaşılan sorunlar arasında;
- Sürenin sınırlılığı,
- Firmaların çok yoğun olarak çalışmaları ve bunun sonucu olarak form doldurma noktasında isteksiz olmaları,
- Firma bilgilerinin gizliliği sebebiyle anket formlarına çok sıcak bakmamaları,
- Genelde anketi dolduracak kişilerin firmadan ziyade sahada bulunmaları nedeniyle ilgili kişilere ulaşılamaması,
- Firmalarda anketi dolduracak personelin yetkili yöneticilerden izinsiz anketi doldurmak istememeleri,
- Firma yetkililerin anket çalışmalarına önyargılı yaklaşmaları,
- Oluşturulan araştırma ölçeğinde yer alan soru sayısının fazla olması nedeniyle katılımcıların anket formunu yanıtlamaya sıcak bakmaması ve bıkkınlık göstermeleri bulunmaktadır.

*C. EVREN VE ÖRNEKLEM*

Çalışmanın evrenini, Türkiye'de bulunan Teknokentler oluşturmaktadır. Çalışmanın örneklemini ise, bu evren içerisinden Ankara ilinde yer alan Gazi Üniversitesi, Hacettepe Üniversitesi, Bilkent Üniversitesi ve ODTÜ olmak üzere dört üniversite oluşturmaktadır. Örneklem belirlenirken, bünyesinde yazılım projesi gerçekleştiren firmaların olup olmadığı dikkate alınmıştır. Örneklem içerisindeki teknokentlerde bulunan yazılım firmaları ve bu firmalarda farklı pozisyonlarda çalışan kişiler katılımcı olarak belirlenmiştir.

*Çizelge 1. Örneklemi oluşturan teknokentler*

| Teknokent | Üniversite Adı | Firma Sayısı | Web Adresi |
|---|---|---|---|
| Gazi Teknopark | Gazi Üniversitesi | 117 | http://www.gaziteknopark.com.tr |
| Hacettepe Teknokent | Hacettepe Üniversitesi | 147 | http://www.hacettepeteknokent.com.tr |
| ODTU Teknokent | Orta Doğu Teknik Üniversitesi | 285 | http://odtuteknokent.com.tr |
| Bilkent Cyberpark | Bilkent Üniversitesi | 225 | http://www.cyberpark.com.tr |

## D. VERİLERİN TOPLANMASI

Araştırma kapsamında, öncelikle daha önce yapılmış gerçek proje verilerini elde etmek amacıyla "Takım Üyeleri (Geliştirici ve Test Elemanı)" ve "Yöneticiler (Analist, Uzman, Takım Lideri, Proje yöneticisi ve Genel Yönetici)" başlıkları altında hazırlanan formlar Google Drive aracılığıyla çevrimiçi sisteme yüklenmiş ve katılımcıların görüşleri doğrultusunda yanıtlamaları için yayına açılmıştır. Bu arada, Türkiye genelinde ulaşılabilinen tüm Teknokent ve birkaç yazılım firmasına yaklaşık 3000 e-mail gönderilmiştir. Ancak tüm bu çabalara rağmen, ilgili formların çevrimiçi olarak geri dönüşleri az sayıda ve eksik bir şekilde gerçekleşmiştir. Tüm bu sebeplerden dolayı, sahaya inilmek zorunda kalınmıştır. Sahada yapılan veri toplama çalışması için, Ankara il sınırları içerisindeki Gazi Üniversitesi, Hacettepe Üniversitesi, Bilkent Üniversitesi ve ODTÜ'de bulunan teknokentlerdeki yazılım geliştirme konusunda proje deneyimine sahip firmalara başvurulmuştur. Kontrol listelerine sahip formlar, yazılım mühendislerine ve özellikle yazılım geliştirme sürecinde bulunan personel ve yöneticilere uygulanmıştır. İhtiyaç duyulan veriler, yaklaşık iki aylık bir süre zarfında toplanmış olup, toplanan veriler ile ilgili bilgiler Çizelge 2'de verilmiştir.

*Çizelge 2. Teknokentlere göre toplanan veriler*

| Teknokent Adı | "Takım Üyesi" Anket Formu Yanıt Sayısı | "Yönetici" Anket Formu Yanıt Sayısı | Toplam |
|---|---|---|---|
| Gazi Üniversitesi Teknopark | 47 | 12 | 59 |
| Hacettepe Teknokent | 92 | 17 | 109 |
| ODTU Teknokent | 72 | 32 | 104 |
| Bilkent Cyberpark | 85 | 16 | 101 |
| Online Sistemden Elde Edilen | 79 | 15 | 94 |
| **Toplam** | 375 | 92 | **467** |

## E. GÜVENİLİRLİK ANALİZİ

Çalışmada kullanılan değişkenlerin güvenilirlik analizi çalışmanın bilimselliği açısından oldukça önemlidir. Yapılan analiz sonucu, çalışmanın "Takım Üyeleri" başlıklı araştırma ölçeğinin güvenirliği 0,949; "Yönetici" başlıklı araştırma ölçeğinin güvenirliği ise 0,927 olarak elde edilmiştir. Yani, her iki pozisyon için elde edilen değer, 0,7'den büyük olup, bu sonuçlar dikkate alındığında değişkenlerin ölçülecek boyutu ölçme noktasında "güvenilir" olduğu söylenebilir.

*Çizelge 3. Güvenilirlik*

| Pozisyon | Cronbach's Alpha | Değişken Sayısı |
|---|---|---|
| Takım Üyeleri | 0,949 | 40 |
| Yönetici | 0,927 | 40 |

# IV. BULGULAR VE TARTIŞMA

Bu bölümde, toplanan veriler üzerinde yapılan istatistiksel analizler sonucunda elde edilen bulgular ve bu bulgular üzerinde yapılan yorumlar yer almaktadır.

## A. FREKANS ANALİZİ

Verilerin değerlendirildiği bu kısımda, anket formunun ilk bölümünde yer alan; proje alanı, pozisyonu, proje süresi, proje bütçesi, personel sayısı; ikinci bölümünde yer alan; proje süresinde, bütçede, personel sayısında ve hedeflerde sapma olup olmadığı ve projenin amacının başarıyla gerçekleşip gerçekleşmediği sorularına ilişkin katılımcıların verdikleri cevapların frekans dağılımlarına yer verilmiştir. Bu verilere ilişkin elde edilen sonuçlar "Takım Üyesi" ve "Yönetici" olmak üzere ayrı ayrı düzenlenmiştir. Araştırma kapsamında, "Projenin alanı" başlığı altında yazılım projeleri içerisinde özellikle Network Uyg., Mobil Uyg. ve İletişim Uygulamaları alanındaki projelerin diğerlerine nazaran daha çok gerçekleştirildiği görülmektedir (Çizelge 4). Bu neticeden, günümüz yazılım endüstrisinde ağ, mobil ve iletişim alanlarının diğer alanlara göre daha fazla önemli görüldüğü veya bu alanlardaki uygulamalara daha fazla ihtiyaç duyulduğu söylenebilir.

*Çizelge 4. Projenin alanı*

| Alan | Takım Üyeleri | | Yönetici | |
|---|---|---|---|---|
| | Frekans | Yüzde | Frekans | Yüzde |
| Mobil Uyg. | 64 | 18,7 | 21 | 17,1 |
| Network Uyg. | 85 | 24,8 | 24 | 19,5 |
| Oyun Uyg. | 13 | 3,8 | 2 | 1,6 |
| İletişim Uyg. | 43 | 12,5 | 23 | 18,7 |
| Web Uyg. | 30 | 8,7 | 7 | 5,7 |
| Eğitim Uyg. | 17 | 5,0 | 5 | 4,1 |
| ERP Uyg. | 9 | 2,6 | 4 | 3,2 |
| Diğer | 82 | 23,9 | 37 | 30,1 |
| **Toplam** | **343** | **100,0** | **123** | **100,0** |

Çizelge 5'te yer alan bulgulara göre; "Pozisyonu" başlığı altında, gerçekleştirilen yazılım projelerinde %82,8'lik bir oranla büyük ölçüde "Geliştirici"ler bulunmaktadır. En düşük düzeyde ise (%8,9), Analist'lerin bulunduğu dikkat çekmektedir. Burada gözden kaçırılmaması gereken bir diğer hususta, normal şartlarda yazılım projelerinde, kodlama kadar test faaliyetlerine de çok önem verilmesine rağmen, elde edilen bulgulara bakıldığında, yazılım firmaları için çok önemli olan "Test Elemanı" pozisyonundaki personel sayısının çok düşük olduğudur. Ayrıca bu durumun, projelerin tahmin edilen süre ve bütçede tamamlanmamasında önemli ölçüde etkili olduğu söylenebilir.

Çizelge 6'da yer alan bulgulara göre ise; "Proje süresi" başlığı altında, katılımcıların en az "24-48" aylık projeler gerçekleştirdiği görülmektedir.

*Çizelge 5.* Pozisyonu

| Pozisyonu | Takım Üyeleri | | Pozisyonu | Yönetici | |
|---|---|---|---|---|---|
| | Frekans | Yüzde | | Frekans | Yüzde |
| Geliştirici | 284 | 82,8 | Analist | 11 | 8,9 |
| Test Elemanı | 59 | 17,2 | Uzman | 22 | 17,9 |
| | | | Takım Lideri | 25 | 20,4 |
| | | | Proje Yöneticisi | 46 | 37,4 |
| | | | Genel Yönetici | 19 | 15,4 |
| **Toplam** | **343** | **100,0** | **Toplam** | **123** | **100,0** |

*Çizelge 6.* Proje süresi

| Proje Süresi | Takım Üyeleri | | Yönetici | |
|---|---|---|---|---|
| | Frekans | Yüzde | Frekans | Yüzde |
| 0 - 6 ay | 69 | 20,1 | 26 | 21,1 |
| 6 - 12 ay | 111 | 32,4 | 33 | 26,8 |
| 12 - 24 ay | 88 | 25,7 | 31 | 25,3 |
| 24 - 48 ay | 28 | 8,2 | 18 | 14,6 |
| 48 ay'dan fazla | 47 | 13,6 | 15 | 12,2 |
| **Toplam** | **343** | **100,0** | **123** | **100,0** |

Çizelge 7'de yer alan bulgulara göre; "Takım üyeleri"nin, hemen hemen her bütçedeki bulunma oranları birbirine yakın elde edilmiştir. "Yönetici" pozisyonundaki katılımcılar için ise, durum biraz farklı olup %36,6 gibi yüksek bir oranla "500000 ve üstü" bütçeye sahip projelerde yer aldıkları görülmektedir.

Çizelge 8'e göre ise, projelerin %70,8'i, personel sayısı 10'dan az olan kişilerle gerçekleştirilmiştir. Elde edilen oranlara bakıldığında, bu oranların her iki örneklem grubu içinde yüksek bir düzeyde (Takım Üyesi için: %70,8 – Yönetici için: %66,7) olduğu görülmektedir. Bu durumun birkaç sebebi olabilir. Örneğin, personel sayısının fazla olması proje maliyetinin artmasına sebep olacağı için yöneticilerin daha ekonomik davranmayı tercih etmiş olabilmeleri, firmaların yazılım sektöründe yetişmiş eleman bulma noktasında sıkıntı çekmiş olabilmeleri veya geliştirilen yazılım projelerinin kapsamının çok büyük olmaması ve buna bağlı olarak da daha fazla personele ihtiyaç duyulmaması gibi.

*Çizelge 7.* Proje bütçesi

| Proje Bütçesi | Takım Üyeleri | | Yönetici | |
|---|---|---|---|---|
| | Frekans | Yüzde | Frekans | Yüzde |
| 50000 TL altı | 73 | 21,3 | 17 | 13,8 |
| 50000 TL - 100000TL | 70 | 20,4 | 17 | 13,8 |
| 100000 TL - 250000 TL | 77 | 22,4 | 23 | 18,7 |
| 250000 TL - 500000 TL | 46 | 13,5 | 21 | 17,1 |
| 500000 TL üstü | 77 | 22,4 | 45 | 36,6 |
| **Toplam** | **343** | **100,0** | **123** | **100,0** |

*Çizelge 8.* Personel sayısı

| Personel Sayısı | Takım Üyeleri | | Yönetici | |
|---|---|---|---|---|
| | Frekans | Yüzde | Frekans | Yüzde |
| 10'dan az | 243 | 70,8 | 82 | 66,7 |
| 10-30 | 67 | 19,5 | 31 | 25,2 |
| 30-50 | 21 | 6,1 | 4 | 3,3 |
| 50-100 | 7 | 2,1 | 3 | 2,4 |
| 100'den fazla | 5 | 1,5 | 3 | 2,4 |
| **Toplam** | **343** | **100,0** | **123** | **100,0** |

Çizelge 9 ve 10'e göre, "Takım Üyesi" personelleri açısından başlatılan projelerin hemen hemen yarısı taahhüt edilen sürede ve bütçede bitirilememiştir. Bu durumun, yazılım projelerinin geleceği açısından endişe verici olduğu söylenebilir. Bu açıdan bakıldığında, test elemanı personelinin eksikliğinin önemi bir kez daha ortaya çıkmaktadır. Ayrıca, "Yönetici"ler açısından projelerin yaklaşık yarısının (47,2) süre aşımına uğramış olduğu, bütçede ise aşımın çok fazla yaşanmadığı görülmektedir (%27,6).

*Çizelge 9. Proje süresinde sapma var mı?*

|  | Takım Üyeleri | | Yönetici | |
|---|---|---|---|---|
|  | Frekans | Yüzde | Frekans | Yüzde |
| Hayır | 175 | 51,0 | 65 | 52,8 |
| Evet | 168 | 49,0 | 58 | 47,2 |
| **Toplam** | **343** | **100,0** | **123** | **100,0** |

*Çizelge 10. Bütçede aşım var mı?*

|  | Takım Üyeleri | | Yönetici | |
|---|---|---|---|---|
|  | Frekans | Yüzde | Frekans | Yüzde |
| Hayır | 175 | 51,0 | 89 | 72,4 |
| Evet | 168 | 49,0 | 34 | 27,6 |
| **Toplam** | **343** | **100,0** | **123** | **100,0** |

Çizelge 11'e göre, geliştirilen yazılım projelerinde kullanılan personel sayısının büyük ölçüde yeterli olduğu görülmektedir. Çizelge 12'ye göre ise, Takım Üyesi ve Yöneticilerin %69,1'i gerçekleştirdikleri projelerin hedeflerinde sapma olmadığını belirtmişlerdir.

*Çizelge 11. Personel sayısında aşım var mı?*

|  | Takım Üyeleri | | Yönetici | |
|---|---|---|---|---|
|  | Frekans | Yüzde | Frekans | Yüzde |
| Hayır | 294 | 85,7 | 99 | 80,5 |
| Evet | 49 | 14,3 | 24 | 19,5 |
| **Toplam** | **343** | **100,0** | **123** | **100,0** |

*Çizelge 12. Hedeflerde sapma var mı?*

|  | Takım Üyeleri | | Yönetici | |
|---|---|---|---|---|
|  | Frekans | Yüzde | Frekans | Yüzde |
| Hayır | 237 | 69,1 | 86 | 69,9 |
| Evet | 106 | 30,9 | 37 | 30,1 |
| **Toplam** | **343** | **100,0** | **123** | **100,0** |

Son olarak, yukarıdaki çizelgelerde görülen proje süresi, bütçesi, personel sayısı ve hedeflerdeki sapmalara rağmen, Çizelge 13'te çok daha olumlu bir tablo ile karşılaşılmaktadır. Çizelge 13'e göre, "Takım Üyesi" ve "Yönetici" pozisyonları açısından bakıldığında, geliştirilen yazılım projelerinin yaklaşık %95'i başarıyla sonuçlanmıştır. Yani, tüm olumsuzluklara rağmen başlatılan yazılım projelerinin büyük çoğunluğu bitirilerek müşteriye teslim edilmiştir.

*Çizelge 13. Projenin amacı başarıyla gerçekleşti mi?*

|  | Takım Üyeleri | | Yönetici | |
|---|---|---|---|---|
|  | Frekans | Yüzde | Frekans | Yüzde |
| Hayır | 20 | 5,8 | 6 | 4,9 |
| Evet | 323 | 94,2 | 117 | 95,1 |
| **Toplam** | **343** | **100,0** | **123** | **100,0** |

Sonuç itibariyle yapılan araştırmaya göre geliştirilen yazılım projelerinin;

% 48,1'inin süresi gecikmiş,

% 38,3'ünün bütçeyi aşmış,

ve % 30,5'inin ise hedefi sapmış olduğu, ancak tüm bunlara rağmen yazılım projelerinin %94,65'inin başarıyla tamamlandığı ortaya çıkmıştır. Bu noktada, Teknokentler'de geliştirilen yazılım projelerinin başarı düzeylerinin oldukça yüksek olduğu söylenebilir.

## B. RİSK FAKTÖRLERİ ARASINDAKİ İLİŞKİLER

Risk faktörleri arasındaki ilişkiler Korelasyon Analizi ile belirlenmiştir. "Takım Üyeleri" ve "Yönetici" başlıklı kontrol listelerinde bulunan risk faktör sayısının fazla olması nedeniyle bu çalışmada sadece en yüksek ilişkili risk faktörleri ve ilişki oranlarına yer verilmiştir (Çizelge 14-15). Her iki çizelge ayrıntılı bir şekilde incelendiğinde, ilişkili risk faktörlerinin konu itibariyle birbirleriyle önemli ölçüde örtüştükleri açıkça görülmektedir.

*Çizelge 14. En yüksek ilişkili risk faktörleri ve ilişki düzeyleri (Takım Üyesi)*

| Risk Faktör No | Risk Faktörünün Adı | İlişkili Olduğu Risk Faktörünün Adı | İlişki Oranı |
|---|---|---|---|
| R6-R8 | Gerçekçi olmayan bir bütçe tahmini yapıldı. | Her birime yeteri derecede bütçe tahsis edilmedi. | 0,752 |
| R8-R10 | Her birime yeteri derecede bütçe tahsis edilmedi. | Kaynak (insan, araç ve materyal) maliyetlerinin belirlenmesinde güçlükler yaşandı. | 0,758 |
| R11-R13 | Yönetici, proje deneyimine sahip değildi. | Yönetici, personel ile sağlıklı bir iletişim kuramadı. | 0,707 |
| R11-R15 | Yönetici, proje deneyimine sahip değildi. | Personel, uzmanlık alanlarına göre görevlendirilmedi. | 0,716 |
| R36-R38 | Ekipman ve sarf malzemesi tesliminde gecikmeler oldu. | Proje süreci içerisinde siyasi istikrarsızlıklar yaşandı. | 0,757 |

*Çizelge 15. En yüksek ilişkili risk faktörleri ve ilişki düzeyleri (Yönetici)*

| Risk Faktör No | Risk Faktörünün Adı | İlişkili Olduğu Risk Faktörünün Adı | İlişki Oranı |
|---|---|---|---|
| R6-R8 | Bütçe değişiklikleri yaşandı. | Kaynak (insan, araç ve materyal) maliyetlerinin belirlenmesinde güçlükler yaşandı. | 0,752 |
| R8-R10 | Kaynak (insan, araç ve materyal) maliyetlerinin belirlenmesinde güçlükler yaşandı. | Proje için yatırım kısıtlamaları mevcuttu. | 0,758 |
| R11-R13 | Personel, sorumluluklarını yerine getirmedi. | Projenin planına göre yönetilmesinde aksaklıklar meydana geldi. | 0,707 |
| R11-R15 | Personel, sorumluluklarını yerine getirmedi. | Takım arası iletişim sağlamada güçlükler çekildi. | 0,716 |
| R36-R38 | Ekipman ve sarf malzemesi tesliminde gecikmeler oldu. | Personel etik ve ahlaki bakımından problemliydi. | 0,757 |

## C. RİSK FAKTÖRLERİNİN POZİSYONLARA GÖRE DAĞILIMI

Burada risk faktörlerine göre, "Geliştirici" ve "Test Elemanı" pozisyonlarında görev alan katılımcıların, hangi risk faktörünü ne düzeyde önemli gördükleri, hangi faktör ile ne oranda karşılaştıkları ve bu faktörleri projeler açısından ne oranda riskli gördükleri ortaya konulmuştur. Burada, her bir risk faktörü için ortalama değerler dikkate alınmıştır.

Çizelge 16'da, risk faktörlerinin "Takım Üyeleri"ne göre "Risk Düzeyini", "Oluşma Sıklıklarını" ve "Riskleri" gösteren dağılımlar verilmiştir. Literatürde risk; risk düzeyi ile riskin oluşma sıklığının çarpımına eşittir. Dolayısıyla, Çizelge 16 yorumlanırken bu durum dikkate alınmıştır. (Açık Gri: İlgili sütundaki en düşük değer; Koyu Gri: İlgili sütundaki en yüksek değer).

*Çizelge 16. Risk faktörlerinin "Takım Üyeleri"ne göre "Risk Düzeyini", "Oluşma Sıklıklarını" ve "Riskleri" gösteren çizelge*

| Risk Grubu | No | Risk Faktörleri | Pozisyonu (Takım Üyesi) | | | | | |
|---|---|---|---|---|---|---|---|---|
| | | | Geliştirici | | | Test Elemanı | | |
| | | | Risk Düzeyi (1-5) | Oluşma Sıklığı (1-5) | Risk | Risk Düzeyi (1-5) | Oluşma Sıklığı (1-5) | Risk |
| Zaman Riskleri | R1 | Görev dağıtımı yanlış yapıldı. | 1,88 | 2,49 | 4,68 | 1,84 | 2,38 | 4,38 |
| | R2 | Donanım dağıtımı yanlış yapıldı. | 1,79 | 2,71 | 4,85 | 1,77 | 2,6 | 4,60 |
| | R3 | Zaman planlaması gerçekçi değildi. | 1,59 | 2,76 | 4,39 | 1,68 | 2,51 | 4,22 |
| | R4 | Projenin yapısında beklenmeyen değişiklikler oldu. | 1,53 | 2,48 | 3,79 | 1,54 | 2,68 | 4,13 |
| | R5 | Proje süreci içerisinde müşteri tarafından zaman kısıtlaması yapıldı. | 2,52 | 2,38 | 6,00 | 2,45 | 2,62 | 6,42 |
| Bütçe (Maliyet) Riskleri | R6 | Gerçekçi olmayan bir bütçe tahmini yapıldı. | 1,9 | 1,77 | 3,36 | 1,75 | 1,7 | 2,98 |
| | R7 | Bütçe planı kaynak durumuna göre güncellenmedi. | 1,77 | 1,99 | 3,52 | 1,8 | 1,96 | 3,53 |
| | R8 | Her birime yeteri derecede bütçe tahsis edilmedi. | 2 | 1,94 | 3,88 | 1,68 | 2,19 | 3,68 |
| | R9 | Bütçe değişiklikleri yaşandı. | 1,76 | 2,07 | 3,64 | 1,77 | 2,15 | 3,81 |
| | R10 | Kaynak (insan, araç ve materyal) maliyetlerinin belirlenmesinde güçlükler yaşandı. | 1,93 | 1,87 | 3,61 | 1,93 | 2 | 3,86 |
| Yönetim Riskleri | R11 | Yönetici, proje deneyimine sahip değildi. | 1,9 | 1,89 | 3,59 | 1,91 | 1,91 | 3,65 |
| | R12 | Kaynak planlaması gerçekçi yapılmadı. | 1,74 | 2,21 | 3,85 | 1,82 | 2,06 | 3,75 |
| | R13 | Yönetici, personel ile sağlıklı bir iletişim kuramadı. | 2,11 | 2,16 | 4,56 | 2,09 | 2,09 | 4,37 |
| | R14 | Proje, plana göre yönetilmedi. | 1,92 | 2,04 | 3,92 | 1,88 | 1,85 | 3,48 |
| | R15 | Personel, uzmanlık alanlarına göre görevlendirilmedi. | 2,08 | 1,9 | 3,95 | 2,11 | 1,74 | 3,67 |
| Teknik Riskler | R16 | Yazılım geliştirme ortamı fiziksel açıdan (güvenlik, çalışma ortamı, radyasyon, malzeme) uygun değildi. | 1,73 | 1,75 | 3,03 | 1,73 | 1,64 | 2,84 |
| | R17 | Kullanılan araç-gereçler yeterli değildi. | 1,66 | 2,22 | 3,69 | 1,68 | 2,17 | 3,65 |
| Planlama ve Program Riskleri | R18 | Proje süreci içerisinde ekipman değişiklikleri meydana geldi. | 1,84 | 2,14 | 3,94 | 1,88 | 2,15 | 4,04 |
| | R19 | Projede personeli zorlayacak yazılım faaliyetleri bulunmaktaydı. | 1,69 | 1,87 | 3,16 | 1,79 | 1,6 | 2,86 |
| | R20 | Gelişmiş teknolojilerden yararlanılamadı. | 1,86 | 1,75 | 3,26 | 1,84 | 1,55 | 2,85 |
| | R21 | Proje kapsamı anlaşılır değildi. | 2,02 | 2,71 | 5,47 | 1,93 | 2,53 | 4,88 |
| | R22 | Yüklenici, kararsız tutumlar sergiledi. | 1,82 | 1,69 | 3,08 | 1,82 | 1,49 | 2,71 |
| | R23 | Proje süreci içerisinde müşterinin öncelikleri değişmekteydi. | 2,27 | 1,58 | 3,59 | 2 | 1,51 | 3,02 |
| | R24 | Malzeme bulmakta güçlükler çekildi. | 2,33 | 2,41 | 5,62 | 2,02 | 2,19 | 4,42 |
| | R25 | Projenin hacmi büyüktü. | 2,7 | 2,36 | 6,37 | 2,61 | 2,43 | 6,34 |
| Sözleşme ve Yasal Riskler | R26 | Sözleşme şartları ağırdı. | 1,98 | 1,92 | 3,80 | 1,98 | 1,68 | 3,33 |
| | R27 | Veri telif hakları ile ilgili problemler yaşandı. | 1,86 | 1,76 | 3,27 | 2 | 1,75 | 3,50 |
| | R28 | Sözleşme metninde değişiklikler meydana geldi. | 1,7 | 1,61 | 2,74 | 1,61 | 1,38 | 2,22 |
| | R29 | Lisanssız yazılım kullanıldı. | 1,62 | 1,85 | 3,00 | 1,54 | 1,66 | 2,56 |
| | R30 | Sözleşme metni, değişen ihtiyaçlara göre güncellenmedi. | 2,04 | 1,74 | 3,55 | 1,89 | 1,58 | 2,99 |
| Personel Riskleri | R31 | Personel sayısı yetersizdi. | 2,23 | 1,85 | 4,13 | 2,39 | 1,81 | 4,33 |
| | R32 | Personel değişiklikleri yaşandı. | 2,35 | 1,94 | 4,56 | 2,48 | 1,91 | 4,74 |
| | R33 | Yönetim, personele mobing (psikolojik baskı) uyguladı. | 2,34 | 1,81 | 4,24 | 2,5 | 1,74 | 4,35 |
| | R34 | Takım üyeleri arasında ekip ruhu yoktu. | 2,38 | 1,73 | 4,12 | 2,63 | 1,62 | 4,26 |
| | R35 | Proje sürecinde personelde sağlık problemleri yaşandı. | 1,97 | 1,68 | 3,31 | 2,05 | 1,58 | 3,24 |
| Diğer Kaynaklı Riskler | R36 | Ekipman ve sarf malzemesi tesliminde gecikmeler oldu. | 1,8 | 1,66 | 2,99 | 1,57 | 1,81 | 2,84 |
| | R37 | Proje, araç ve tesis bakımından yetersizdi. | 1,69 | 1,35 | 2,28 | 1,75 | 1,45 | 2,54 |
| | R38 | Proje süreci içerisinde siyasi istikrarsızlıklar yaşandı. | 1,65 | 1,22 | 2,01 | 1,61 | 1,11 | 1,79 |
| | R39 | Proje süreci içerisinde doğal felaketler yaşandı. | 1,61 | 1,51 | 2,43 | 1,68 | 1,53 | 2,57 |
| | R40 | Personeller arasında kültürel farklılıklar vardı. | 1,67 | 1,56 | 2,61 | 1,77 | 1,57 | 2,78 |

Risk faktörleri "Takım Üyeleri" açısından incelendiğinde, "Geliştirici" ve "Test Elemanı" pozisyonundaki katılımcılara göre, *"Zaman Riskleri"* grubunda yer alan "Proje süreci içerisinde müşteri tarafından zaman kısıtlaması yapıldı." riskinin en yüksek düzeyde; "Projenin yapısında beklenmeyen değişiklikler oldu." riskinin ise en düşük düzeydeki olduğu ortaya çıkmıştır.

*"Bütçe (Maliyet) Riskleri"* grubuna bakıldığında, "Geliştirici" pozisyonundaki katılımcılar için en yüksek düzeydeki riskin "Her birime yeteri derecede bütçe tahsis edilmedi."; "Test Elemanı" pozisyonundaki katılımcılar için ise, "Kaynak (insan, araç ve materyal) maliyetlerinin belirlenmesinde güçlükler yaşandı." olduğu; her iki pozisyon için en düşük düzeydeki risk ise, "Gerçekçi olmayan bir bütçe tahmini yapıldı." faktörü olduğu ortaya çıkmıştır.

*"Yönetim Riskleri"* grubuna bakıldığında, "Geliştirici" ve "Test Elemanı" pozisyonundaki katılımcılar için en yüksek düzeydeki riskin "Yönetici, personel ile sağlıklı bir iletişim kuramadı." olduğu görülmektedir. Yine aynı gruptaki "Geliştirici" pozisyonundaki katılımcılar için en düşük düzeydeki riskin, "Yönetici, proje deneyimine sahip değildi."; "Test Elemanı" pozisyonundaki katılımcılar için ise, "Proje, plana göre yönetilmedi." olduğu ortaya çıkmıştır.

*"Teknik Riskler"* grubuna bakıldığında, "Geliştirici" ve "Test Elemanı" pozisyonundaki katılımcılar için en yüksek düzeydeki riskin "Proje süreci içerisinde ekipman değişiklikleri meydana geldi."; en düşük düzeydeki riskin ise, "Yazılım geliştirme ortamı fiziksel açıdan (güvenlik, çalışma ortamı, radyasyon, malzeme) uygun değildi." olduğu ortaya çıkmıştır.

*"Planlama ve Program Riskleri"* grubunda, "Geliştirici" ve "Test Elemanı" pozisyonundaki katılımcılar için, en yüksek düzeydeki riskin "Projenin hacmi büyüktü.", en düşük düzeydeki riskin ise "Yüklenici, kararsız tutumlar sergiledi." olduğu ortaya çıkmıştır.

*"Sözleşme ve Yasal Riskler"* grubuna bakıldığında, "Geliştirici" pozisyonundaki katılımcılar için en yüksek düzeydeki riskin, "Sözleşme şartları ağırdı."; "Test Elemanı" pozisyonundaki katılımcılar için ise, "Veri telif hakları ile ilgili problemler yaşandı." olduğu; yine aynı gruptaki "Geliştirici" ve "Test Elemanı" pozisyonundaki katılımcılar için en düşük düzeydeki riskin ise, "Sözleşme metninde değişiklikler meydana geldi." olduğu ortaya çıkmıştır.

*"Personel Riskleri"* grubunda, "Geliştirici" ve "Test Elemanı" pozisyonundaki katılımcılar için en yüksek düzeydeki riskin, "Personel değişiklikleri yaşandı."; en düşük düzeydeki riskin ise "Proje sürecinde personelde sağlık problemleri yaşandı." olduğu ortaya çıkmıştır.

*"Diğer Kaynaklı Riskler"* grubuna bakıldığında ise, "Geliştirici" ve "Test Elemanı" pozisyonundaki katılımcılar için en yüksek düzeydeki riskin, "Ekipman ve sarf malzemesi tesliminde gecikmeler oldu."; en düşük düzeydeki riskin ise, "Proje süreci içerisinde siyasi istikrarsızlıklar yaşandı." olduğu ortaya çıkmıştır.

Çizelge 17'de ise, Risk faktörlerinin "Yönetici"lere göre "Risk Düzeyini", "Oluşma Sıklıklarını" ve "Riskleri" gösteren dağılımlar verilmiştir. Belirtilen değerler, Çizelge 16'da olduğu gibi yine ortalama değerlerdir.

*Çizelge 17. Risk faktörlerinin "Yönetici"lere göre "Risk Düzeyini", "Oluşma Sıklıklarını" ve "Riskleri" gösteren çizelge*

| Risk Grubu | No | Risk Faktörleri | Analist | | | Uzman | | | Takım Lideri | | | Proje Yöneticisi | | | Genel Yönetici | | |
|---|---|---|---|---|---|---|---|---|---|---|---|---|---|---|---|---|---|
| | | | Risk Düzeyi (1-5) | Oluşma Sıklığı (1-5) | Risk | Risk Düzeyi (1-5) | Oluşma Sıklığı (1-5) | Risk | Risk Düzeyi (1-5) | Oluşma Sıklığı (1-5) | Risk | Risk Düzeyi (1-5) | Oluşma Sıklığı (1-5) | Risk | Risk Düzeyi (1-5) | Oluşma Sıklığı (1-5) | Risk |
| Zaman Riskleri | R1 | Zamanlama planının hazırlanmasında zorluklar yaşandı. | 2,56 | 2,80 | 7,17 | 1,94 | 2,05 | 3,98 | 2,70 | 2,81 | 7,59 | 2,14 | 1,91 | 4,09 | 1,65 | 1,89 | 3,12 |
| | R2 | Donanım dağıtımında problemler meydana geldi. | 2,22 | 3,00 | 6,66 | 2,06 | 2,26 | 4,66 | 2,30 | 2,81 | 6,46 | 1,91 | 2,33 | 4,45 | 1,53 | 2,11 | 3,23 |
| | R3 | Proje teslim süresinde gecikme oldu. | 1,89 | 2,50 | 4,73 | 1,67 | 2,16 | 3,61 | 1,78 | 2,62 | 4,66 | 1,53 | 2,09 | 3,20 | 1,29 | 2,16 | 2,79 |
| | R4 | Hazırlanan ve onaylanan zamanlama planın uygulanmasında zorluklar yaşandı. | 1,44 | 3,20 | 4,61 | 1,56 | 2,16 | 3,37 | 1,91 | 2,95 | 5,63 | 1,40 | 2,33 | 3,26 | 1,18 | 2,37 | 2,80 |
| | R5 | Proje yapısında beklenmeyen değişiklikler oldu. | 2,89 | 2,70 | 7,80 | 2,00 | 2,11 | 4,22 | 2,57 | 2,81 | 7,22 | 2,05 | 2,35 | 4,82 | 2,00 | 2,11 | 4,22 |
| Bütçe Riskleri | R6 | Bütçe değişiklikleri yaşandı. | 2,11 | 2,30 | 4,85 | 1,72 | 1,79 | 3,08 | 2,04 | 2,43 | 4,96 | 1,77 | 1,77 | 3,13 | 1,65 | 1,89 | 3,12 |
| | R7 | Proje, hedeflenen ve onaylanan bütçe ile tamamlanamadı. | 2,00 | 2,90 | 5,80 | 1,67 | 2,11 | 3,52 | 1,96 | 2,38 | 4,66 | 1,63 | 2,16 | 3,52 | 1,47 | 1,84 | 2,70 |
| | R8 | Kaynak (insan, araç ve materyal) maliyetlerinin belirlenmesinde güçlükler yaşandı. | 2,33 | 2,70 | 6,29 | 2,06 | 2,00 | 4,12 | 2,30 | 2,38 | 5,47 | 1,95 | 1,74 | 3,39 | 1,53 | 1,74 | 2,66 |
| | R9 | Gerçekleşen ile planlanan maliyet arasında fark oluştu. | 2,56 | 2,20 | 5,63 | 2,06 | 2,05 | 4,22 | 2,43 | 2,29 | 5,56 | 1,72 | 2,09 | 3,59 | 1,47 | 1,89 | 2,78 |
| | R10 | Proje için yatırım kısıtlamaları mevcuttu. | 2,11 | 2,20 | 4,64 | 1,72 | 1,95 | 3,35 | 2,61 | 2,29 | 5,98 | 2,02 | 1,77 | 3,58 | 1,82 | 1,79 | 3,26 |
| Yönetim Riskl. | R11 | Personel, sorumluluklarını yerine getirmedi. | 1,78 | 2,70 | 4,81 | 1,61 | 2,00 | 3,22 | 1,87 | 2,19 | 4,10 | 1,81 | 1,65 | 2,99 | 1,65 | 1,74 | 2,87 |
| **Teknik Riskler** | R12 | Kaynaklar, etkili bir şekilde kullanılamadı. | 2,00 | 3,20 | 6,40 | 1,67 | 2,11 | 3,52 | 1,91 | 2,67 | 5,10 | 1,72 | 1,88 | 3,23 | 1,47 | 1,79 | 2,63 |
| | R13 | Projenin planına göre yönetilmesinde aksaklıklar meydana geldi. | 2,33 | 3,10 | 7,22 | 1,72 | 2,32 | 3,99 | 1,87 | 2,81 | 5,25 | 1,77 | 1,72 | 3,04 | 1,41 | 1,95 | 2,75 |
| | R14 | Önceliklerin doğru bir şekilde sıralanmasında zorlular yaşandı. | 2,00 | 2,40 | 4,80 | 1,72 | 2,11 | 3,63 | 1,83 | 1,76 | 3,22 | 1,70 | 1,81 | 3,08 | 1,35 | 1,37 | 1,85 |
| | R15 | Takım arası iletişim sağlamada güçlükler çekildi. | 3,00 | 2,20 | 6,60 | 1,94 | 2,16 | 4,19 | 2,09 | 1,86 | 3,89 | 1,81 | 1,72 | 3,11 | 1,65 | 1,74 | 2,87 |
| | R16 | Yazılım geliştirme ortamı fiziksel açıdan (güvenlik, çalışma ortamı, radyasyon, malzeme) uygun değildi. | 2,00 | 1,90 | 3,80 | 1,39 | 1,42 | 1,97 | 1,61 | 1,48 | 2,38 | 1,42 | 1,30 | 1,85 | 1,18 | 1,63 | 1,92 |
| | R17 | Kullanılan araç-gereçler yeterli değildi. | 1,44 | 2,20 | 3,17 | 1,28 | 2,42 | 3,10 | 1,26 | 2,33 | 2,94 | 1,19 | 2,14 | 2,55 | 1,00 | 2,47 | 2,47 |
| | R18 | Proje süreci içerisinde ekipman değişiklikleri meydana geldi. | 1,78 | 2,20 | 3,92 | 1,56 | 2,21 | 3,45 | 1,35 | 2,14 | 2,89 | 1,35 | 1,72 | 2,32 | 1,06 | 2,16 | 2,29 |
| | R19 | Projede personeli zorlayacak yazılım faaliyetleri bulunmaktaydı. | 1,56 | 2,10 | 3,28 | 1,39 | 1,58 | 2,20 | 1,26 | 1,52 | 1,92 | 1,14 | 1,51 | 1,72 | 1,12 | 1,16 | 1,30 |
| | R20 | Gelişmiş teknolojilerden yararlanılamadı. | 1,67 | 1,80 | 3,01 | 1,50 | 1,47 | 2,21 | 1,61 | 1,62 | 2,61 | 1,49 | 1,40 | 2,09 | 1,35 | 1,26 | 1,70 |
| **Planlama ve Program Riskleri** | R21 | Proje kapsamı anlaşılır değildi. | 2,44 | 3,20 | 7,81 | 1,89 | 2,58 | 4,88 | 1,87 | 2,95 | 5,52 | 1,86 | 2,70 | 5,02 | 1,18 | 3,00 | 3,54 |
| | R22 | Yüklenici, kararsız tutumlar sergiledi. | 2,56 | 2,30 | 5,89 | 1,78 | 1,42 | 2,53 | 1,78 | 1,76 | 3,13 | 1,60 | 1,35 | 2,16 | 1,29 | 1,11 | 1,43 |
| | R23 | Proje süreci içerisinde müşterinin öncelikleri değişmekteydi. | 2,56 | 1,70 | 4,35 | 2,22 | 1,47 | 3,26 | 2,04 | 1,71 | 3,49 | 2,05 | 1,26 | 2,58 | 1,71 | 1,21 | 2,07 |
| | R24 | Malzeme bulmakta güçlükler çekildi. | 3,22 | 2,40 | 7,73 | 2,33 | 2,42 | 5,64 | 2,13 | 2,71 | 5,77 | 2,09 | 2,14 | 4,47 | 2,18 | 2,00 | 4,36 |
| | R25 | Projenin hacmi büyüktü. | 3,11 | 2,50 | 7,78 | 2,39 | 2,42 | 5,78 | 2,83 | 2,76 | 7,81 | 2,79 | 2,05 | 5,72 | 2,41 | 1,84 | 4,43 |
| **Sözleşme ve Yasal Riskler** | R26 | Sözleşme şartları ağırdı. | 2,11 | 1,40 | 2,95 | 1,83 | 1,68 | 3,07 | 2,09 | 2,19 | 4,58 | 1,98 | 1,37 | 2,71 | 1,29 | 1,68 | 2,17 |
| | R27 | Veri telif hakları ile ilgili problemler yaşandı. | 1,56 | 2,30 | 3,59 | 1,83 | 1,42 | 2,60 | 1,96 | 1,33 | 2,61 | 1,72 | 1,42 | 2,44 | 1,41 | 1,21 | 1,71 |
| | R28 | Sözleşme metninde değişiklikler meydana geldi. | 1,67 | 1,20 | 2,00 | 1,56 | 1,32 | 2,06 | 1,61 | 1,52 | 2,45 | 1,49 | 1,49 | 2,22 | 1,29 | 1,26 | 1,63 |
| | R29 | Lisanssız yazılım kullanıldı. | 1,67 | 2,70 | 4,51 | 1,56 | 1,42 | 2,22 | 1,74 | 1,90 | 3,31 | 1,37 | 1,56 | 2,14 | 1,18 | 1,37 | 1,62 |

|  |  |  |  |  |  |  |  |  |  |  |  |  |  |  |  |
|---|---|---|---|---|---|---|---|---|---|---|---|---|---|---|---|
| **Personel Riskleri** | R30 | Paydaşlar yapılan sözleşmeye uymadı. | 1,78 | 2,40 | 4,27 | 1,67 | 1,32 | 2,20 | 2,13 | 1,57 | 3,34 | 1,70 | 1,53 | 2,60 | 1,35 | 1,47 | 1,98 |
|  | R31 | Personel sayısı yetersizdi. | 2,44 | 2,00 | 4,88 | 1,83 | 2,00 | 3,66 | 2,65 | 1,90 | 5,04 | 1,91 | 1,77 | 3,38 | 1,94 | 1,53 | 2,97 |
|  | R32 | Personel değişiklikleri yaşandı. | 1,89 | 1,70 | 3,21 | 2,00 | 1,58 | 3,16 | 2,65 | 1,62 | 4,29 | 1,79 | 1,40 | 2,51 | 1,76 | 1,37 | 2,41 |
|  | R33 | Personelin eğitim düzeyi yeterli değildi. | 2,78 | 1,50 | 4,17 | 2,33 | 1,58 | 3,68 | 2,74 | 1,48 | 4,06 | 2,42 | 1,40 | 3,39 | 2,06 | 1,58 | 3,25 |
|  | R34 | Proje sürecinde personelde sağlık problemleri yaşandı. | 2,67 | 2,30 | 6,14 | 2,33 | 1,68 | 3,91 | 2,70 | 1,67 | 4,51 | 2,02 | 1,56 | 3,15 | 1,94 | 1,32 | 2,56 |
|  | R35 | Takım üyeleri arasında ekip ruhu yoktu. | 1,67 | 1,90 | 3,17 | 2,11 | 1,58 | 3,33 | 1,91 | 1,48 | 2,83 | 1,79 | 1,47 | 2,63 | 1,47 | 1,42 | 2,09 |
| **Diğer Kaynaklı Riskler** | R36 | Ekipman ve sarf malzemesi tesliminde gecikmeler oldu. | 1,89 | 1,50 | 2,84 | 1,17 | 1,00 | 1,17 | 1,70 | 1,33 | 2,26 | 1,42 | 1,12 | 1,59 | 1,18 | 1,42 | 1,68 |
|  | R37 | Proje, araç ve tesis bakımından yetersizdi. | 1,67 | 1,90 | 3,17 | 1,22 | 1,79 | 2,18 | 1,39 | 1,81 | 2,52 | 1,37 | 1,19 | 1,63 | 1,18 | 1,47 | 1,73 |
|  | R38 | Personel etik ve ahlaki bakımından problemliydi. | 1,56 | 1,30 | 2,03 | 1,33 | 1,63 | 2,17 | 1,43 | 1,62 | 2,32 | 1,30 | 1,21 | 1,57 | 1,00 | 1,26 | 1,26 |
|  | R39 | Proje süreci içerisinde siyasi istikrarsızlıklar yaşandı. | 1,44 | 1,60 | 2,30 | 1,17 | 1,21 | 1,42 | 1,17 | 1,19 | 1,39 | 1,23 | 1,14 | 1,40 | 1,00 | 1,16 | 1,16 |
|  | R40 | Proje süreci içerisinde doğal felaketler yaşandı. | 1,56 | 1,00 | 1,56 | 1,11 | 1,00 | 1,11 | 1,48 | 1,14 | 1,69 | 1,30 | 1,12 | 1,46 | 1,24 | 1,05 | 1,30 |

Çizelge 17'de, pozisyon sayısının fazla olması nedeniyle sadece en çok dikkat çeken maddeler yorumlanmıştır. Burada özellikle, "Analist", "Proje Yöneticisi" ve "Genel Yönetici" pozisyonundaki katılımcılar için en yüksek düzeydeki riskin, "Proje yapısında beklenmeyen değişiklikler oldu." olduğu görülmüştür.

*"Bütçe Riskleri"* başlığı altında muhtelif pozisyonda farklı sonuçlar elde edilmiştir. Ancak, özellikle "Analist" başlığı altındaki katılımcılar için yüksek bir oranla, en yüksek düzeydeki riskin "Kaynak (insan, araç ve materyal) maliyetlerinin belirlenmesinde güçlükler yaşandı." olduğu ortaya çıkmıştır.

*"Yönetim Riskleri"* başlığı altında muhtelif pozisyonda farklı sonuçlar elde edilmiştir. Ancak, yine "Analist" başlığı altındaki katılımcılar için yüksek bir oranla, en yüksek düzeydeki riskin "Projenin planına göre yönetilmesinde aksaklıklar meydana geldi." olduğu dikkat çekmektedir.

*"Teknik Riskler"* başlığı altında "Takım Lideri", "Proje Yöneticisi" ve "Genel Yönetici" pozisyonundaki katılımcılar için en yüksek düzeydeki riskin, "Kullanılan araç-gereçler yeterli değildi."; en düşük riskin ise, "Projede personeli zorlayacak yazılım faaliyetleri bulunmaktaydı." olduğu görülmüştür. Ayrıca, "Analist" ve "Uzman" pozisyonundaki katılımcılar için en yüksek risk, "Proje süreci içerisinde ekipman değişiklikleri meydana geldi." olduğu görülmüştür.

*"Planlama ve Program Riskleri"* başlığı altında "Uzman", "Takım Lideri", "Proje Yöneticisi" ve "Genel Yönetici" pozisyonundaki personeller için en yüksek riskin, "Projenin hacmi büyüktü."; en düşük riskin ise "Yüklenici, kararsız tutumlar sergiledi." olduğu dikkat çekmektedir.

*"Sözleşme ve Yasal Riskler"* başlığı altında "Uzman", "Takım Lideri", "Proje Yöneticisi" ve "Genel Yönetici" pozisyonundaki personeller için en yüksek risk, "Sözleşme şartları ağırdı."; en düşük riskler ise, "Sözleşme metninde değişiklikler meydana geldi." ve "Lisanssız yazılım kullanıldı." olarak görülmektedir.

*"Personel Riskleri"* başlığı altında muhtelif pozisyonda farklı sonuçlar elde edilmiştir. Ancak, "Analist" pozisyonundaki personeller için "Proje sürecinde personelde sağlık problemleri yaşandı." en yüksek risk olduğu görülmektedir.

*"Diğer Kaynaklı Riskler"* başlığı altında ise, tüm pozisyonlardaki personeller için en yüksek riskin, "Proje, araç ve tesis bakımından yetersizdi."; en düşük ise, "Proje süreci içerisinde siyasi istikrarsızlıklar yaşandı." ve "Proje süreci içerisinde doğal felaketler yaşandı." olduğu ortaya çıkmıştır.

Son olarak, Çizelge 16-17'de yer alan risk faktörleri, en yüksek "Risk" puanına göre öncelik sırasına konulmaktadır. Daha sonra, oluşturulan öncelikli risklerin yönetilmesi için bir plan hazırlanıp yönetimin onayına sunulmaktadır. Böylece, risk analizi süreci tamamlanmış olmaktadır.

# V. SONUÇ VE ÖNERİLER

Yapılan çalışmada, geliştirilen anket formlarında bulunan kontrol listeleri aracılığıyla, Teknokent'lerde geliştirilen yazılım projelerinde karşılaşılan risk faktörleri belirlenip analiz edilmiş ve bu projelerin başarı düzeyleri ortaya konulmuştur. Elde edilen sonuçlara göre, geliştirilen yazılım projelerinin; %48,1'inin süresi gecikmiş, %38,3'ünün bütçeyi aşmış ve %30,5'inin ise hedefi sapmış olduğu, ancak bunun yanısıra tüm bu olumsuz sonuçlara rağmen başlatılan yazılım projelerinin yaklaşık %95 gibi çok yüksek bir oranla bitirildiği ve müşterilere teslim edildiği görülmüştür. Buradan, geliştirilen yazılım projelerinin özellikle personel sayısı, proje süresi ve bütçesi gibi konularda ciddi risk altında olduğu açıkça ortaya çıkmaktadır. Bu durum, Teknokent'ler bünyesindeki yazılım firma yöneticilerinin veya proje yetkililerinin yazılım bütçesi, personel sayısı ve süresi ile ilgili konularda iyi analiz ve tahmin yapamadıklarının, projenin geleceği ile ilgili isabetli bir öngörüde bulunamadıklarının bir delili olarak gösterilebilir.

Yazılım projelerinde, "Takım Üyesi" personellerine göre en büyük risk olarak "Proje süreci içerisinde müşteri tarafından zaman kısıtlaması yapıldı." ve "Projenin hacmi büyüktü." görüldüğü; "Yönetici" personellerine göre ise, "Zamanlama planının hazırlanmasında zorluklar yaşandı.", "Proje yapısında beklenmeyen değişiklikler oldu.", "Kaynak (insan, araç ve materyal) maliyetlerinin belirlenmesinde güçlükler yaşandı.", "Projenin planına göre yönetilmesinde aksaklıklar meydana geldi.", "Proje kapsamı anlaşılır değildi." ve "Projenin hacmi büyüktü." olarak görüldüğü ortaya çıkmıştır.

Dikkat çekilmesi gereken bir diğer husus da, proje sürecinde yazılım test elemanı personeli kullanan firmaların çok az sayıda olmasıdır. Bu noktada, firmaların en az "Geliştirici-Kodlayıcı" kadar, "Test Elemanı" personellerini de bünyelerinde bulundurmaları, onlardan da istifade etmeleri ve bu konuya önem vermeleri gerektiği söylenebilir. Aksi takdirde projeler, her ne kadar başarılı bir şekilde müşterilere teslim edilse de, en azından teslim sonrasında bakım-onarım faaliyetleri sürecinde ciddi risklerle karşı karşıya kalınması kaçınılmaz olacaktır.

Ayrıca proje yöneticilerinin, projelerin daha sağlıklı ilerlemesi açısından yazılım ve yazılım proje yönetimi konularında daha fazla eğitimli olmaları gerekmektedir. Günümüz yazılım endüstrisinde, zorlu rekabet koşullarında ayakta kalmak isteyen firmalar için yazılım projelerinin öneminin gün geçtikçe arttığı düşünüldüğünde, yapılan çalışmanın yazılım proje yöneticileri ve bu alanda hizmet veren firmalar açısından önemli ve anlamlı olacağı düşünülmektedir.

# VI. KAYNAKLAR